\documentclass[12pt,epsf]{article}
\usepackage{axodraw}
\usepackage{amsmath}
\usepackage{amssymb}
\usepackage{amsfonts}
\usepackage{bbm} 
\usepackage{graphics}
\input epsf

\begin{document}
\baselineskip 24pt

\newcommand{\be}{\begin{equation}}
\newcommand{\ee}{\end{equation}}

\newcommand{\bea}{\begin{eqnarray}}
\newcommand{\eea}{\end{eqnarray}}
\newcommand{\nn}{\nonumber}
\newcommand{\eq}{\ref}
\newcommand{\ol}{\overline}

\def\<{\left\langle}
\def\>{\right\rangle}

%%%%%%%%

\newcommand{\sheptitle}
{Dirac Neutrinos and Hybrid Inflation from String Theory} 

\newcommand{\shepauthor}
{S. Antusch, O. J. Eyton-Williams and S. F. King}

\newcommand{\shepaddress}
{School of Physics and Astronomy, University of Southampton, \\
        Southampton, SO17 1BJ, U.K.\vspace{1cm}}
\vspace{0.25in}

%%%%%%%%%%%%%%%%%%%%%%%%%%%%%%%%%%%%%%%%%%%%%%%%%%%%%%%%%%%%%%%%%%%%%%%%%%%%%%%
%                                  ABSTRACT
%%%%%%%%%%%%%%%%%%%%%%%%%%%%%%%%%%%%%%%%%%%%%%%%%%%%%%%%%%%%%%%%%%%%%%%%%%%%%%%
\newcommand{\shepabstract}
  { \vspace{-1cm}We consider a possible scenario for the generation of Dirac
  neutrino masses motivated by 
  Type I string theory. 
  The smallness of the neutrino Yukawa couplings is
  explained by an anisotropic compactification with one compactification 
  radius larger than the others. In addition to this
  we utilise small Yukawa couplings to develop strong links between the origin of neutrino masses and
  the physics driving inflation. 
  We construct a minimal model which 
  simultaneously accommodates small Dirac neutrino masses leading to 
  bi-large lepton mixing as well as an inflationary solution to 
  the strong CP and to the $\mu$ problem. 
  }

\begin{titlepage}
\begin{flushright}
%hep-ph/0505xxx \\
SHEP-0514 \\
\end{flushright}
\begin{center}
{\large{\bf \sheptitle}}
\\ \shepauthor \\ \mbox{} \\ {\it \shepaddress} \\
{\bf Abstract} \bigskip \end{center} \setcounter{page}{0}
\shepabstract
\begin{flushleft}
%\today
\end{flushleft}

\vskip 0.1in
\noindent

\end{titlepage}

\newpage

\section{Introduction} \label{sec:Introduction}

The evidence of neutrino masses is the first clear signal of physics beyond
the Standard Model (SM). The most popular explanation of neutrino masses and of
their smallness is the well known see-saw mechanism, where heavy SM-singlet 
right-handed neutrinos are introduced. Another possibility, even more minimal
than the see-saw mechanism in the sense that it has less free parameters, 
is that neutrinos have pure Dirac masses 
$m^\nu_{\mathrm{LR}} = Y_\nu v_\mathrm{u}$, generated from a Yukawa coupling 
\begin{eqnarray}
(Y_\nu)_{ij} \overline{L}_i H_\mathrm{u} \nu^{\mathrm{c}}_{\mathrm{R} j} \;.
\end{eqnarray}       
Obviously, generating neutrino masses of order $0.1$ eV requires Yukawa
couplings $(Y_\nu)_{ij} \sim 10^{-12}$. This required smallness is the main 
objection against Dirac neutrinos - and we are going to address it in this
letter. 
Existing explanations of this smallness utilise for instance  
right-handed singlet neutrinos propagating in the bulk, or allow only highly 
suppressed effective operators, e.g.\ by linking the smallness of neutrino 
masses to supersymmetry breaking \cite{DiracNus,Alternatives}.
Heterotic string constructions can also lead to Dirac neutrinos 
 and in some classes of Heterotic orbifolds Dirac neutrino
masses may even be more favoured than 
the see-saw mechanism \cite{Giedt:2005vx}.

Small couplings are also required for the inflationary
solution to the strong CP and to the $\mu$ problem of the MSSM, proposed in \cite{Bastero-Gil:1997vn,Eyton-Williams:2004bm}. The $\mu$-term arises from 
a superpotential term 
\begin{eqnarray}
\lambda \phi H_\mathrm{u} H_\mathrm{d}  
\end{eqnarray}  
within the model of inflation proposed in \cite{Eyton-Williams:2004bm}. 
The vev $\<\phi\>$ of the inflaton after inflation generates $\mu=\lambda
\<\phi\>$ 
and furthermore breaks Peccei-Quinn symmetry solving the strong CP
problem. Satisfying the constraints on the axion decay constant and the scale of $\mu$ requires 
$\<\phi\> = 10^{13}$ GeV and a small coupling $\lambda$ of order $10^{-10}$.  

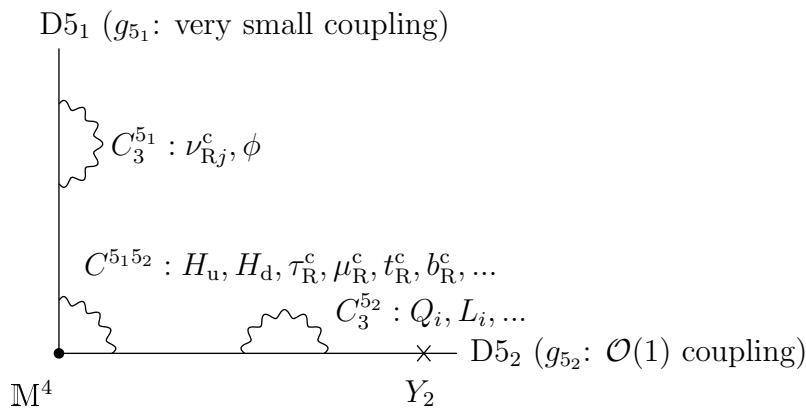
\begin{figure}[h!]
  \begin{picture}(300,135)(0,-10)
                                %D branes
     \Line(100,10)(250,10)
     \Text(256,10)[l]{D$5_2$ ($g_{5_2}$: $\mathcal{O}(1)$ coupling)}
     \Line(100,10)(100,125)
     \Text(93,134)[l]{D$5_1$ ($g_{5_1}$: very small coupling)}
                                %String States
    \PhotonArc(100,10)(20,0,90){1.5}{4.5}
    \PhotonArc(185,10)(15,0,180){1.5}{6.5}
    \PhotonArc(100,89)(15,-90,90){1.5}{6.5}
                                %Twisted Moduli
    \Line(235,13)(240,7)
    \Line(240,13)(235,7)
    \Text(237,-5)[c]{$Y_2$}
                                %String Labels
    \Text(110,43)[l]{$C^{5_1 5_2}: H_\mathrm{u}, H_\mathrm{d},
    \tau^{\mathrm{c}}_{\mathrm{R}}, \mu^{\mathrm{c}}_{\mathrm{R}},
    t^{\mathrm{c}}_{\mathrm{R}}, b^{\mathrm{c}}_{\mathrm{R}}, ...$}
    \Text(205,26)[l]{$C^{5_2}_3: Q_{i}, L_{i}, ...$}
    \Text(120,88)[l]{$C^{5_1}_3: \nu^{\mathrm{c}}_{\mathrm{R} j}, \phi$}
    \Vertex(100,10){2}
    \Text(90,-5)[c]{$\mathbbm{M}^4$}
  \end{picture} \caption{\small 
  Graphical illustration of the origin of small couplings in our scenario. 
  Chiral matter corresponds to open string states $C$ with ends confined either to 
  one of the three orthogonal stacks of D5-branes or to the stack of 
  space-filling D9-branes. 
  While the gauge coupling associated with the stack of branes 
  D$5_2$ is $\mathcal {O}(1)$, the gauge
  coupling associated with D$5_1$ is $\mathcal {O}(10^{-10})$. 
  Fields assigned to states $C_3^{5_1}$ can only participate in interactions 
  with small couplings. 
  The stacks of branes overlap in Minkowski space $\mathbbm{M}^4$, but are 
  orthogonal in the compactified dimensions. 
  $Y_2$ is a twisted modulus 
  localised within the extra dimensions, but 
  free to move in Minkowski space.
 } \label{fig:smallcouplings}
  \end{figure}

From the above discussion, it is clear that very small Yukawa couplings are a
prerequisite for both Dirac neutrinos and for the inflationary solution to the
strong CP and $\mu$ problem.  It should also be noted that both the right-handed
neutrinos $\nu_{\mathrm{R i}}$ and the inflaton $\phi$ are
special in the sense that they should have only small couplings
to ordinary matter and that they are effectively SM-singlets. In 
\cite{Eyton-Williams:2005bg} it has been shown how such small couplings might
arise in the context of Type I string theory. 
The details are reviewed in Appendix \ref{sec:Couplings}. 
The mechanism has the 
simple graphical illustration shown in Fig.~\ref{fig:smallcouplings}. Matter corresponds to open string
states with ends confined either to one of the three orthogonal stacks of
D5-branes or to the stack of space-filling D9-branes. With a  
compactification radius for one of the D5-branes\footnote{The branes wrap the compact dimensions hence they can be associated with their radii.} (D5$_1$ in Fig.~\ref{fig:smallcouplings}) being larger than the
compactification radius for the others it can be shown that the gauge and Yukawa couplings associated with this brane are small.   Fields corresponding to open strings 
confined to D5$_2$ branes with both ends can only participate in interactions 
with small couplings and it is some of these fields that we will identify with
the right-handed neutrinos. The D5$_2$ branes wrap with a smaller radius and
hence have a gauge coupling large enough to be associated with the Standard Model.  As a result the Standard Model gauge group must arise from this stack of branes.

In this letter, we construct a minimal string-motivated model which 
simultaneously accommodates small Dirac neutrino masses leading to bi-large 
lepton mixing as well as the inflationary solution \cite{Eyton-Williams:2004bm} to 
the strong CP and to the $\mu$ problem.

By string motivated, we mean that we take some restrictions from string theory,
i.e.\ so-called string selection rules, and construct a field theory model
consistent with these rules. We do not claim that an embedding of our particular
model in string theory
is possible. In fact, we do expect that there will emerge additional constraints if one
attempts to embed our scenario in string theory. Nevertheless, we find it useful
to identify attractive routes for explaining the smallness of Dirac
neutrino masses within the framework of string theory from a bottom-up perspective.
      
The layout of the rest of the paper is as follows.  Section \ref{sec:Restrictions} provides a brief discussion of the string framework we consider in this paper to clarify its use in model building.  This is followed by the main body of the paper, section \ref{sec:TheModel}, in which we detail the model, its construction, the masses and MNS mixings.  Having discussed the lepton sector section \ref{sec:InflationModel} considers the inflation model that could be realised within the same framework. The conclusions are to be found in section \ref{sec:Conclusions} and are followed by two appendices.  In Appendix \ref{sec:Couplings} we present a more thorough account of the string framework and we clarify our lepton mixing conventions in Appendix \ref{conventions}.

\section{Restrictions and benefits from string selection rules} \label{sec:Restrictions}

Let us briefly state how the scenario outlined in the introduction and 
illustrated in Fig.~\ref{fig:smallcouplings} restricts model building.  
For the details we refer the reader to Appendix \ref{sec:Couplings}. 
In particular we would like to draw the reader's attention to 
Eq.~(\ref{eq:WGen}) from which we extract the following terms to be 
utilised in the model:
\begin{align}
 W= g_{5_1}C^{5_1}_3 C^{5_1 5_2} C^{5_1 5_2} 
 + g_{5_2} C^{5_2}_3 C^{5_1 5_2} C^{5_1 5_2}\; . \label{eq:Wintro}
\end{align}
$C^{5_1}_3$, $C^{5_2}_3$ and $C^{5_1 5_2}$ are low energy excitations of strings: 
charged chiral superfields.  The superscripts denote the branes 
which the strings end on and terms with different subscripts 
transform differently under the gauge group associated with the brane.
The string construction involves assigning fields to low energy string 
excitations (the $C$ terms) and showing that only the 
gauge invariant operators 
appearing in Eq.~(\ref{eq:WGen}) can be written down.  
For the purpose of our model we will make use of the fact that the only
 couplings of $C^{5_1}_3$ and 
$C^{5_2}_3$ to intersection states $C^{5_1 5_2}$ appearing in
Eq.~(\ref{eq:WGen}) are those of Eq.~(\ref{eq:Wintro}). 
The $C$ terms can 
have more than one field assigned to them and the fields can only 
transform under the gauge groups of the branes to which the $C$ field 
attaches.  For example the $C^{5_1 5_2}$ term has fields that transform 
under the gauge groups of the D5$_1$ and D5$_2$ branes.  If the operator does 
not appear in Eq.~(\ref{eq:WGen}) or cannot transform under gauge groups 
required by the field theory then we say such an operator is forbidden 
by the string selection rules.  For more details about the rules and 
the explicit construction we refer the reader to \cite{Eyton-Williams:2005bg} 
and merely quote the results here.
The gauge couplings $g_{5_1}, g_{5_2}, g_{5_3}$ and $g_{9}$ on the branes are related to the 
Planck scale $M_p$ and the string scale $M_*$ by
\begin{align}
  g_{5_1}g_{5_2}g_{5_3}g_{9}=32\pi^2\left(\frac{M_*}{M_p}\right)^2. 
  \label{eq:gaugereln}
\end{align}
Here we shall choose the couplings 
$g_{5_2}=\sqrt{\frac{4\pi}{24}}$ (to give $\alpha_{\mbox{\tiny
GUT}}=1/24$, consistent with gauge coupling unification),
$g_{5_3}=g_9=2$ and $g_{5_1}=10^{-10}$, which results in the string scale 
$M_*= 10^{13}$ GeV.\footnote{Gauge coupling unification at the string scale, $M_*= 10^{13}$, where the gauge coupling on the brane, $g_{5_2}$, is specified, is possible due to power law
running. The requirement of gauge coupling unification would constrain 
the matter content and mass spectrum of complete models - 
which is however beyond the scope of this letter. In the presence of twisted
moduli, this constraint would be modified.  } 
The gauge couplings on the branes are related to the extra-dimensional radii as 
$
  g_{5_i}^2=2\pi \lambda_{\mathrm{I}} / (R_i^2 M_*^2) 
 , g_{9}^2=2\pi \lambda_{\mathrm{I}}/(R_1^2 R_2^2 R_3^2 M_*^6) 
$, 
where $\lambda_{\mathrm{I}}$ is the ten dimensional dilaton that governs the strength 
of string interactions. 

Thus, the geometry of the compactification determines
the couplings of the theory. Furthermore,  
assigning the superfields of a model to open string excitations restricts the
possible superpotential couplings to the ones given in 
Eq.~(\ref{eq:WGen}).

\section{Dirac neutrinos} \label{sec:TheModel}
With the right-handed neutrino fields $\nu^{\mathrm{c}}_{\mathrm{R} i}$ assigned to string states $C_3^{5_1}$ 
and with the gauge coupling $g_{5_1}$ on the branes D5$_1$ being of order
$10^{-10}$ (the choice $g_{5_1} = \mathcal{O}(10^{-10})$ is motivated by the hybrid
inflation model which we will discuss in Sec.~\ref{sec:InflationModel}) 
light neutrinos of Dirac-type naturally emerge - all couplings to 
$\nu^{\mathrm{c}}_{\mathrm{R} i}$ are suppressed by, at least, $g_{5_1}$. On the other hand,
right-handed charged leptons assigned to string states $C_3^{5_2}$ corresponds
to the larger charged lepton masses since 
the gauge coupling $g_{5_2}$ on the branes D5$_2$ is of order $1$.

\begin{figure}[h!]
  \begin{picture}(300,120)(0,0)
    \SetWidth{1} 
    % Left Propagators
    \ArrowLine(40,20)(70,60)
    \Text(30,10)[l]{$L_{2}$}
    \ArrowLine(40,100)(70,60)
    \Text(30,110)[l]{$H_\mathrm{d}$}
    \Text(45,60)[l]{$g_{5_2}$}
    \Text(171,60)[l]{$g_{5_2}$}
    \Text(112,50)[l]{$g_{5_2}$} 
    \Text(291,60)[l]{$g_{5_2}$}    
    % Right Propagators
    \ArrowLine(190,100)(160,60)
    \Text(200,110)[r]{$\mu^{\mathrm{c}}_{\mathrm{R}}$}
    \ArrowLine(190,20)(160,60)
    \Text(200,10)[r]{$A$}
    % Internal Line
    \ArrowLine(115,60)(70,60)
    \ArrowLine(115,60)(160,60)
    \Text(92,50)[]{$\chi_{A}$}
    \Text(142,50)[]{$\overline{\chi}_{A}$}
    % Mass term
    \ArrowLine(115,60)(115,80)
    % The Cross
    \Line(112,83)(118,77)
    \Line(118,83)(112,77)
    \Text(115,92)[]{$\<\psi\>$}
    % To the next diagram
    \LongArrow(210,60)(230,60)
    % Left Propagators
    \ArrowLine(250,20)(280,60)
    \Text(240,10)[l]{$L_2$}
    \ArrowLine(250,100)(280,60)
    \Text(240,110)[l]{$H_\mathrm{d}$}
    % Right Propagators
    \ArrowLine(310,100)(280,60)
    \Text(320,110)[r]{$\mu^{\mathrm{c}}_{\mathrm{R}}$}
    \ArrowLine(310,20)(280,60)
    \Text(320,10)[r]{$A$}
  \end{picture}
  \caption{\small Froggatt Nielsen supergraphs leading to 
  the muon mass. Higher-dimensional FN diagrams can generate NNLO
  Yukawa couplings, e.g.\ for realising the electron mass. 
  } \label{fig:FNmuon}
\end{figure}
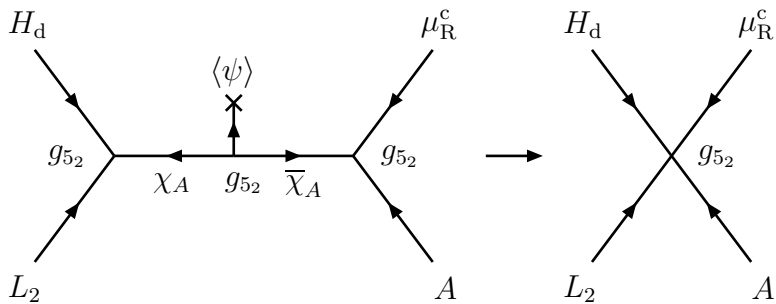

Let us now consider the generation of lepton masses in more detail. 
With 
$H_\mathrm{u}$ and the right-handed tau $\tau^{\mathrm{c}}_{\mathrm{R}}$ assigned to
intersection states $C^{5_1 5_2}$ and the lepton doublet assigned to a string
state $C_3^{5_2}$, we see from Eq.~(\ref{eq:Wintro}) that a renormalizable
Yukawa coupling $y_\tau \sim g_{5_2}=\mathcal{O}(1)$ 
\begin{eqnarray}
g_{5_2} C_3^{5_2} C^{5_1 5_2}C^{5_1 5_2}  &\rightarrow& 
g_{5_2} L_i  H_\mathrm{d} \tau^{\mathrm{c}}_{\mathrm{R}} \quad \; \mbox{($\mathcal {O}(1)$ coupling
$g_{5_2}$)} \label{Eq:mtau}
\end{eqnarray}
is allowed by the string selection rules. 
As we have seen, Yukawa couplings in this string motivated setup cannot be chosen freely, but are
fixed by the values of gauge couplings. The $\mathcal {O}(1)$-coupling to
$\tau^{\mathrm{c}}_{\mathrm{R}}$ of Eq.~(\ref{Eq:mtau}) can be consistent with low energy
experimental data for appropriately chosen large $\tan \beta$.\footnote{Note that if 3rd
family quark Yukawa couplings also stem from analogous renormalizable couplings,
3rd family gauge-Yukawa unification $y_\tau \sim y_b \sim y_t \sim g_{5_2}$ 
for the SM fermions holds, up to order 1 coefficients, even
though they are not unified in an irrep of a unified gauge group. 
We will not address the details of the quark 
sector in this letter. } 
However, the Yukawa couplings leading e.g.\ to the mass $m_\mu$ of the muon, though allowed by the string selection rules, 
do not stem from an analogous term to Eq.~(\ref{Eq:mtau}) since they will be forbidden by the symmetries of the model discussed below. Our strategy is to
obtain it from a non-renormalizable operator generated via a
supersymmetric Froggatt-Nielsen (FN) \cite{Froggatt:1978nt} 
mechanism, for example
\begin{eqnarray}
\frac{g_{5_2}}{\< \psi \>} \< A \> L_2  H_\mathrm{d} \mu^{\mathrm{c}}_{\mathrm{R}}\; .
\end{eqnarray} 
This is represented in Fig.~\ref{fig:FNmuon}. Such Froggatt-Nielsen supergraphs
are allowed by the string selection rules if the messenger fields are
assigned to intersection states and if the masses for the messenger fields stem
from the vev $\< \psi \>$ of the scalar component of an additional field $\psi$, assigned to
a string state $C_3^{5_2}$. The flavon field $A$ has to be assigned to
a string state $C_3^{5_2}$ as well and the muon mass from this operator 
is suppressed by $\< A \>/\< \psi \>$. Generating the muon mass of the right order requires $\< A \> /\<\psi\> = 
\mathcal {O}(10^{-3})$.\footnote{Note that we do not have to specify the values of
the vevs at this stage. We should however keep in mind that that $\<\psi\>$,
i.e.\ the mass of the messenger fields in the FN mechanism, should be
sufficiently below the masses of the winding modes of the messenger fields 
(in our scenario $\approx
n 10^8$ GeV, see Appendix \ref{sec:Couplings}), so that the effects of the extra dimensions are under control.} 
We have ignored the electron mass here, which can be generated by
higher-dimensional operators in a straightforward way.  

Let us now consider the neutrino sector: 
since atmospheric neutrino oscillations
suggest a neutrino mass scale $m_3 \approx \sqrt{\Delta m^2_{31}} \approx 0.05$ 
eV \cite{SuperK}, a  
Yukawa coupling $(Y_\nu)_{ij}\approx 10^{-12} - 10^{-13}$ 
is required for
generating the largest neutrino mass eigenvalue $m_3$. 
Thus, a renormalizable Yukawa coupling $(Y_\nu)_{ij} = 
g_{5_1} \sim 10^{-10}$ would be already in the right range,
 but still somewhat too large. In addition to $m_3 \approx 0.05$ eV the 
 experimental results for 
$\Delta m^2_{21}$ require $m_2  \approx \sqrt{\Delta m^2_{21}} \approx 0.01$ eV, 
which also requires a suppression of about 
$10^{-3}$ compared to $g_{5_1}$.

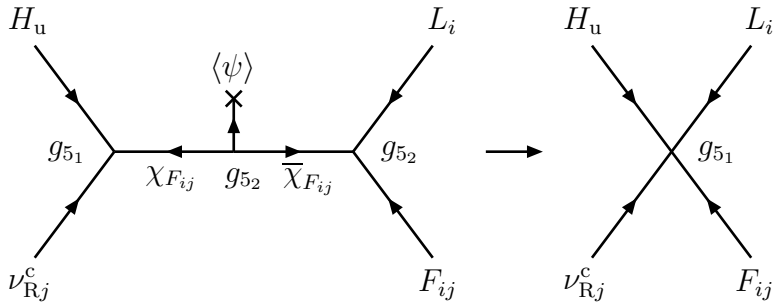
\begin{figure}[h!]
  \begin{picture}(300,120)(0,0)
    \SetWidth{1} 
    % Left Propagators
    \ArrowLine(40,20)(70,60)
    \Text(30,10)[l]{$\nu^{\mathrm{c}}_{\mathrm{R} j}$}
    \ArrowLine(40,100)(70,60)
    \Text(30,110)[l]{$H_\mathrm{u}$}    
    \Text(45,60)[l]{$g_{5_1}$}
    \Text(171,60)[l]{$g_{5_2}$}
    \Text(112,50)[l]{$g_{5_2}$}  
    \Text(291,60)[l]{$g_{5_1}$}    

    % Right Propagators
    \ArrowLine(190,100)(160,60)
    \Text(200,110)[r]{$L_i$}
    \ArrowLine(190,20)(160,60)
    \Text(200,10)[r]{$F_{ij}$}
    % Internal Line
    \ArrowLine(115,60)(70,60)
    \ArrowLine(115,60)(160,60)
    \Text(92,50)[]{$\chi_{F_{ij}}$}
    \Text(144,50)[]{$\overline{\chi}_{F_{ij}}$}
    % Mass term
    \ArrowLine(115,60)(115,80)
    % The Cross
    \Line(112,83)(118,77)
    \Line(118,83)(112,77)
    \Text(115,92)[]{$\<\psi\>$}
    % To the next diagram
    \LongArrow(210,60)(230,60)
    % Left Propagators
    \ArrowLine(250,20)(280,60)
    \Text(240,10)[l]{$\nu^{\mathrm{c}}_{\mathrm{R} j}$}
    \ArrowLine(250,100)(280,60)
    \Text(240,110)[l]{$H_\mathrm{u}$}
    % Right Propagators
    \ArrowLine(310,100)(280,60)
    \Text(320,110)[r]{$L_i$}
    \ArrowLine(310,20)(280,60)
    \Text(320,10)[r]{$F_{ij}$}
  \end{picture}
  \caption{\small Froggatt Nielsen supergraphs leading to 
  neutrino Yukawa couplings which are slightly suppressed compared to the 
  already small tree-level value $g_{5_1} = 10^{-10}$. 
  $F_{ij}$     % \in \{a,b,c,e,f,A\}$ 
  are flavon superfields and $\chi_{F_{ij}},\bar{\chi}_{F_{ij}}$
  are corresponding messenger superfields. 
  } \label{fig:FN}
\end{figure}

In fact, we see that the string selection rules forbid the renormalizable tree
level Yukawa couplings involving $\nu^{\mathrm{c}}_{\mathrm{R} j}$ and $L_i$. However,  
as in the charged lepton sector for $m_\mu$, we can rely on a FN mechanism   
(cf.\ Fig.~\ref{fig:FN}) for obtaining the neutrino Yukawa couplings which
can then also have the desired additional suppression. 
The neutrino Yukawa couplings can then stem from the leading order effective operators
\begin{eqnarray}
\frac{g_{5_1}}{\< \psi \>} \<F_{ij}\> L_i  H_\mathrm{u} 
\nu^{\mathrm{c}}_{\mathrm{R} j} \; .
\end{eqnarray}
From Fig.~\ref{fig:FN} and Eq.~(\ref{eq:Wintro}) we can determine appropriate string
assignments of the flavons and the messenger superfields and see that the messenger fields have to be assigned to intersection states $C^{5_1 5_2}$.  The flavons are found to be both intersection and single brane states. 
In the neutrino sector, suppression factors similar to the one for 
the muon mass,   
$\<F_{ij}\>/\< \psi \> = \mathcal{O}(10^{-2}) ...\, \mathcal{O}(10^{-3})$, are 
needed.

In summary, the Yukawa matrices of neutrinos 
and charged leptons could stems from LO and NLO operators 
of the following form (up to $\mathcal {O}(1)$-factors):
\begin{eqnarray}
Y_\nu &:& \frac{g_{5_1}}{\< \psi \>} \<F_{ij}\> L_i  H_\mathrm{u} \nu^{\mathrm{c}}_{\mathrm{R} j} \; ,\\
Y_e &:& g_{5_2} L_3  H_\mathrm{d} \tau^{\mathrm{c}}_{\mathrm{R}} + 
\frac{g_{5_2}}{\< \psi \>} \< A \> L_2  H_\mathrm{d} \mu^{\mathrm{c}}_{\mathrm{R}}\; .
\end{eqnarray}  

The fact that the renormalizable Yukawa couplings $(Y_{\nu})_{ij}$ are forbidden helps in three ways. Firstly, as already noted, it can lead to the
desired additional suppression compared to the already small gauge coupling 
$g_{5_1} =\mathcal{O} (10^{-10})$. Secondly, if we use flavour symmetries for determining the
structure of $Y_e$ and $Y_\nu$, we find that for renormalisable operators 
large off-diagonal elements in $Y_\nu$ would come along with identical 
large off-diagonal elements in $Y_e$ - making it
difficult to construct the desired large neutrino mixings. As we will see below,
large lepton mixing can easily be achieved if the renormalisable coupling is forbidden. Thirdly,
there is only a very mild mass hierarchy $m_3 / m_2 \lesssim 5$ for $m_2$ and
$m_3$ in the neutrino
sector, compared 3rd and 2nd generation masses of quarks and charged leptons. 
This can be explained by the Yukawa couplings relevant for the neutrino 
masses $m_3$ and $m_2$ being  
generated at the same (or similar) order in the FN mechanism - 
which can be a consequence of the forbidden renormalizable term.

Majorana masses for the right-handed neutrinos are not allowed by a
renormalisable term due to string selection rules. Higher-dimensional operators
for Majorana masses are suppressed by $g_{5_1}^n \sim (10^{-10})^n$, with
$n\ge 2$. For obtaining pure Dirac masses, we can impose in addition a global
U(1)$_{\mathrm{B-L}}$ which forbids Majorana masses for $\nu^{\mathrm{c}}_{\mathrm{R} i}$ to
all orders.    

Let us consider an explicit example with a hierarchical neutrino mass spectrum
and how bi-large neutrino mixing can be realised within this scheme. We will
assume that the lightest neutrino mass $m_1$ is approximately zero. Effectively, this means
that only two right-handed neutrinos are required or equivalently that $\nu_{R 1}$'s Yukawa couplings are generated at higher order and hence heavily surpressed.  For our analysis we treat its couplings as being zero. 
For generating an appropriate set of operators which leads to the observed 
bi-large lepton mixing, we will use a $\mathbb{Z}_3$-symmetry and the
U(1)$_\mathrm{R}$-symmetry (which will be broken to matter parity) 
in addition to the SM
gauge group SU(3)$_\mathrm{C} \times$ SU(2)$_{\mathrm{L}}\times$ U(1)$_\mathrm{Y}$
which is a subgroup of the gauge group $G$, a copy of which is associated with 
the branes D5$_2$.\footnote{The U(1)$_\mathrm{R}$-symmetry is 
broken down to its $\mathbb{Z}_4$ subgroup by the appearance of 
gauginos' soft masses. 
This $\mathbb{Z}_4$ is in turn broken to its $\mathbb{Z}_2$ subgroup, 
matter-parity, when $A$ obtains its vev.  We require that this breaking 
takes place before the end of inflation to avoid the domain wall problem.}
The field content and the corresponding 
 charge assignments of our minimal model is listed in table 
\ref{tab:Matterandflavons} and table \ref{tab:Messengers}, which contain the
matter superfields and flavon superfields $F_{ij} \in \{a,b,c,e,f\}$ and $A$.  They also contain the superfields of the messenger sector $\chi_{F_{ij}} \in \{\chi_a,\chi_e,\chi_A\}$,
respectively. The superpotential then contains the following renormalisable 
terms:  
\begin{align}
  W_{ren.} = \; \:&\: g_{5_2}  L_3 H_\mathrm{d}\tau^{\mathrm{c}}_{\mathrm{R}} + g_{5_2}  L_2 H_\mathrm{d}\chi_A + g_{5_2} \chi_A \psi \bar{\chi}_A + g_{5_2} \bar{\chi}_A \mu^{\mathrm{c}}_{\mathrm{R}} A\nonumber \\ 
  +&\:  g_{5_1} \chi_aH_\mathrm{u} \nu^{\mathrm{c}}_{\mathrm{R} 2}  +  g_{5_2} \chi_a \psi \bar{\chi}_a +  g_{5_2} L_1\bar{\chi}_a   a +  g_{5_2} L_2\bar{\chi}_a  b + g_{5_2} L_3\bar{\chi}_a  c \nonumber \\ 
  +&\:  g_{5_1}\chi_e H_\mathrm{u} \nu^{\mathrm{c}}_{\mathrm{R} 3}  +  g_{5_2} \chi_e \psi \bar{\chi}_e +  g_{5_2} L_2 \bar{\chi}_e  e +  g_{5_2} L_3\bar{\chi}_e  f \nonumber\\ 
  +&\:  g_{5_1} \phi H_\mathrm{u} H_\mathrm{d} + g_{5_1} \phi N^2 +  g_{5_2} Q_3
  H_\mathrm{u} t^{\mathrm{c}}_{\mathrm{R}} + g_{5_2}  Q_3 H_\mathrm{d}
  b^{\mathrm{c}}_{\mathrm{R}}. \label{eq:Wren}
\end{align}
From this superpotential, assuming that the flavons develop vevs,  
the FN diagrams of Fig.~\ref{fig:FNmuon} and Fig.~\ref{fig:FN} 
lead to the mass matrices of the neutrinos and charged leptons\footnote{It is
easy to check that all the messenger fields get heavy when the flavons and
$\psi$ obtain their vevs.} 
\begin{equation}\label{eq:NeutrinoMass}
  m^\nu_{\mathrm{LR}} \sim \left(
  \begin{array}{ccc}
    0 & \<a\> & 0 \\
    0 & \<b\> & \<e\> \\
    0 & \<c\> & \<f\> \\
  \end{array} 
  \right) \cdot \frac{ 
  g_{5_1} \<H_\mathrm{u}\>}{\<\psi\>} 
\;,
\quad
  m^{\mathrm{E}}_{\mathrm{LR}} \sim \left(
  \begin{array}{ccc}
    0 & 0 & 0 \\
    0 & \<A\>/\<\psi\> & 0 \\
    0 & 0 & {\mathcal O}(1) \\
  \end{array}
\right)\cdot 
g_{5_2}  \<H_\mathrm{d}\> \; .
\end{equation}

\begin{table}
  \begin{center}
  \begin{tabular}{|c|c|c|c|c|c|c|c|}
    \hline
    & SU(3)$_\mathrm{C}$  &  SU$(2)_\mathrm{L}$ & U$(1)_\mathrm{Y}$ & U$(1)_\mathrm{R}$ & $\mathbb{Z}_3$ & String State \\
    \hline
    $Q_3$ & \bf{3} & \bf{2} & 1/6 & -1/2 & 1 & $C_3^{5_2}$\\
    \hline 
    $t^{\mathrm{c}}_{\mathrm{R}}$ & $\bar{\mathbf{3}}$ & $\mathbf{1}$ & -2/3 & -1/2 & 1 & $C^{5_1 5_2}$\\
    \hline
    $b^{\mathrm{c}}_{\mathrm{R}}$ & $\bar{\mathbf{3}}$ & $\mathbf{1}$ & 1/3 & -1/2 & 1 & $C^{5_1 5_2}$\\
    \hline
    $H_\mathrm{u}$ & $\mathbf{1}$ & \textbf{2} & 1/2 & 1 & 1  & $C^{5_1 5_2}$\\
    \hline
    $H_\mathrm{d}$ & $\mathbf{1}$ & \textbf{2} & -1/2 & 1 & 1  & $C^{5_1 5_2}$\\
    \hline
    $\nu^{\mathrm{c}}_{\mathrm{R} 2}$ & $\mathbf{1}$ & $\mathbf{1}$ & 0 & -3/2 & 0  & $C_3^{5_1}$\\
    \hline
    $\nu^{\mathrm{c}}_{\mathrm{R} 3}$ & $\mathbf{1}$ & $\mathbf{1}$ & 0 & -7/2 & 1  & $C_3^{5_1}$\\
    \hline
    $L_1$ & $\mathbf{1}$ & \textbf{2} & -1/2 & -1/2 & 2  & $C_3^{5_2}$\\
    \hline
    $L_2$ & $\mathbf{1}$ & \textbf{2} & -1/2 & -3/2 & 1  & $C_3^{5_2}$\\
    \hline
    $L_3$ & $\mathbf{1}$ & \textbf{2} & -1/2 & -5/2 & 1  & $C_3^{5_2}$\\
    \hline
    $\mu^{\mathrm{c}}_{\mathrm{R}}$ & $\mathbf{1}$ & $\mathbf{1}$ & 1 & 3/2 & 1  & $C^{5_1 5_2}$\\
    \hline
    $\tau^{\mathrm{c}}_{\mathrm{R}}$ & $\mathbf{1}$ & $\mathbf{1}$ & 1 & 7/2 & 1  & $C^{5_1 5_2}$\\
    \hline
    $\phi$ & $\mathbf{1}$ & $\mathbf{1}$ & 0 & 0 & 1 & $C_3^{5_1}$ \\
    \hline
    $N$ & $\mathbf{1}$ & $\mathbf{1}$ & 0 & 1 & 1  & $C^{5_1 5_2}$\\
    \hline
    $A$ & $\mathbf{1}$ & $\mathbf{1}$ & 0 & 1 & 0 & $C_3^{5_2}$\\
    \hline
    $a$ & $\mathbf{1}$ & $\mathbf{1}$ & 0 & 3 & 0 & $C^{5_1 5_2}$ \\
    \hline
    $b$ & $\mathbf{1}$ & $\mathbf{1}$ & 0 & 4 & 1 & $C^{5_1 5_2}$ \\
    \hline
    $c$ & $\mathbf{1}$ & $\mathbf{1}$ & 0 & 5 & 1 & $C^{5_1 5_2}$ \\
    \hline
    $e$ & $\mathbf{1}$ & $\mathbf{1}$ & 0 & 6 & 0 & $C^{5_1 5_2}$ \\
    \hline
    $f$ & $\mathbf{1}$ & $\mathbf{1}$ & 0 & 7 & 0 & $C^{5_1 5_2}$ \\
    \hline     
    $\psi$ & $\mathbf{1}$ &  $\mathbf{1}$ & 0 & 0 & 0 & $C_3^{5_2}$\\
      \hline 
  \end{tabular}

  \parbox{200pt}{
  \caption{\label{tab:Matterandflavons} Matter fields and flavons}} 
      \end{center}
\end{table}
\begin{table}
  \begin{center}
    \begin{tabular}{|c|c|c|c|c|c|c|c|}
      \hline 
       &SU(3)$_\mathrm{C}$  &  SU$(2)_\mathrm{L}$ & U$(1)_\mathrm{Y}$ & U$(1)_\mathrm{R}$ & $\mathbb{Z}_3$ & String State \\
      \hline
       $\chi_A$ & $\mathbf{1}$ & \textbf{1} & 1 & 5/2 & 1 & $C^{5_1 5_2}$\\
      \hline
       $\overline{\chi}_A$ & $\mathbf{1}$ & \textbf{1} & -1 & -1/2 & 2 & $C^{5_1 5_2}$\\
      \hline
       $\chi_e$  & $\mathbf{1}$& \textbf{2} & -1/2 & 9/2 & 1 & $C^{5_1 5_2}$\\
      \hline
       $\overline{\chi}_e$ & $\mathbf{1}$& \textbf{2} & 1/2 & -5/2 & 2 & $C^{5_1 5_2}$\\
      \hline
       $\chi_a$ & $\mathbf{1}$ & \textbf{2} & -1/2 & 5/2 & 2 & $C^{5_1 5_2}$\\
      \hline
       $\overline{\chi}_a$ & $\mathbf{1}$ & \textbf{2} & 1/2 & -1/2 & 1 & $C^{5_1 5_2}$\\
      \hline
    \end{tabular} 
    
    \parbox{200pt}{
      \caption{\label{tab:Messengers} Messenger fields}}
  \end{center}
\end{table}

Let us now discuss neutrino masses and lepton mixing with mass
matrices of the structure given in Eq.~(\ref{eq:NeutrinoMass}). 
We assume the sequential dominance of the Dirac neutrino Yukawa couplings
\begin{equation}
\<e\>,\<f\> \gg \<a\>,\<b\>,\<c\>,
\end{equation}
where much larger means
here larger by about a factor 5.  Then, in this approximation, the
couplings to $\nu^{\mathrm{c}}_{\mathrm{R}3}$ will lead to the
neutrino mass eigenvalue $m_3\simeq 0.05$ eV and the couplings to
$\nu^{\mathrm{c}}_{\mathrm{R}2}$ to the smaller mass eigenvalue $m_2
\simeq 0.01$ eV. Clearly, since $m_{\mathrm{LR}}^\mathrm{E}$ is
diagonal, the lepton mixing matrix $U_{\mathrm{MNS}} = R_{23} U_{13}
R_{12}$ will be entirely given by the diagonalization matrix, i.e.\
$U_{\mathrm{MNS}} = U^\dagger_{\nu_\mathrm{L}}$ with
$\mbox{diag}(m_1,m_2,m_3) = U_{\nu_\mathrm{L}}
\,m^\nu_{\mathrm{LR}}\,U^\dagger_{\nu_\mathrm{R}}$. Thus we find for
the MNS mixings (assuming real flavon vevs for
simplicity):\footnote{Since we assume a form of sequential dominance
(SD) \cite{SequentialD} for $Y_\nu$, it is not surprising that the
formulae for the mixing angles are very similar to the ones for
see-saw neutrino masses under the assumption of SD.}
\begin{eqnarray}
\tan (\theta_{23}) &\approx& \frac{\<e\>}{\<f\>} \; ,\\ \label{eq:tantheta23}
\tan (\theta_{12}) &\approx& \frac{\<a\>}{c_{23}\<b\> - s_{23}\<c\>} \; ,\\ 
\theta_{13} &\approx& 0 \vphantom{\frac{\<e\>}{\<f\>}}\; ,
\end{eqnarray}
with $m_3$ and $m_2$ given by
\begin{eqnarray}
m_3 &\approx& \sqrt{\<e\>^2+\<f\>^2}\: \frac{g_{5_1}v_\mathrm{u}}{\<\psi\>} \;\:=\:\;
\frac{\<e\>}{s_{23}} \: \frac{g_{5_1}v_\mathrm{u}}{\<\psi\>}\; ,\\
m_2 &\approx& \sqrt{\<a\>^2 + (c_{23}\<b\> - s_{23}\<c\>)^2} \: \frac{g_{5_1}v_\mathrm{u}}{\<\psi\>} \;\:=\:\;
\frac{\<a\>}{s_{12}} \: \frac{g_{5_1}v_\mathrm{u}}{\<\psi\>} \; ,\\
m_1 &\approx& 0 \vphantom{\frac{\<e\>}{\<f\>}}\;. \label{eq:m1}
\end{eqnarray}  
We see that obtaining nearly maximal atmospheric mixing $\theta_{23}$ requires
$\<e\> \approx \<f\>$. Obtaining a large (but non-maximal) solar mixing $\theta_{12}$
 requires $\sqrt{2} \<a\> \sim \<b\> - \<c\>$. Clearly, the neutrino data fixes
 $\<e\>/\<\psi\>,\<f\>/\<\psi\>$ (from $m_3$ and $\theta_{23} \approx \pi/4$), $\<a\>/\<\psi\>$ (from
 $m_2$) and the combination $(\<b\> - \<c\>)/\<\psi\>$ from consistency with experimental
 data for $\theta_{12}$ (currently $\theta_{12} \approx 30^\circ$). We have
 neglected effects from the RG evolution of the neutrino parameters at this
 stage.

Finally, we remark that although the right-handed neutrinos 
$\nu^{\mathrm{c}}_{\mathrm{R} i}$ as well as the
inflaton $\phi$ are SM-singlets, we shall have in mind that at some stage, 
they transform in a representation of the gauge group $G$, a copy of which 
is associated with each of the three stacks of D-branes.\footnote{This is one
point which 
distinguishes our approach from approaches where right-handed neutrinos are
gauge singlets and propagate in the bulk.} 
$G$ is just (spontaneously) broken completely on D$5_1$, whereas on 
D$5_2$ the SM gauge group $G_{321} \subset G$ is unbroken. To demonstrate this,
let us consider two additional U(1) symmetries, U(1)$_{5_1}$ on D$5_1$ and 
U(1)$_{5_2}$ on D$5_2$, and assign charges to the SM-singlet fields
$\phi,\nu^{\mathrm{c}}_{\mathrm{R} i},N$ and $\psi$. 
First, giving $H_\mathrm{u}$
and $H_\mathrm{d}$ U(1)$_{5_1}$-charge $1$, we see that $\phi$ has charge $-2$
and thus $N$ has charge $1$. From the FN diagram in Fig.~\ref{fig:FN} we can
determine the charges of the messenger fields if we assign a U(1)$_{5_1}$-charge $q$
to the right-handed neutrinos $\nu^{\mathrm{c}}_{\mathrm{R} i}$ and finally the flavons
$F_{ij}$, which are intersection states $C^{5_1 5_2}$, end up with charge 
$-(q+1)$. 
Note that only fields which
are assigned to string states $C^{5_1}_3$ and $C^{5_1 5_2}$ can be charged under
U(1)$_{5_1}$ and only fields $C^{5_2}_3$ and $C^{5_1 5_2}$ can 
be charged under U(1)$_{5_2}$. 
Similarly for the $C^{5_2}_3$ state $\psi$, 
from the FN diagram in Fig.~\ref{fig:FN} we see how giving
it a U(1)$_{5_2}$-charge $p$ determines e.g.\ the charge of 
$\bar \chi_{F_{ij}}$
to be $-p$ and the charge of the flavons $F_{ij}$ to be $p$. It is easy to see
that this charge assignment can be extended consistently to all the fields of
the model.

\section{A brief review of the hybrid inflation model}\label{sec:InflationModel}

We will now review the inflation model \cite{Eyton-Williams:2004bm} where the
required small couplings might also originate from the string motivated scenario
outlined here \cite{Eyton-Williams:2005bg}. In fact, the requirement of 
very small couplings - and their possible common solution - might provide a link between the 
origin of (Dirac) neutrino masses and the physics driving inflation.

The starting point of the field theory model is the following part of the
 superpotential Eq.~(\ref{eq:Wren}) relevant to inflation (with
 $\lambda \sim \kappa \sim g_{5_1} = 10^{-10}$) and the corresponding soft terms:
\begin{align}\label{eq:NewW}
  W_{inf.}=\lambda \phi H_\mathrm{u} H_\mathrm{d} + \kappa \phi N^2.
\end{align}
\begin{align}\label{eq:NewV}
  V_{soft} = & \;V(0)+\lambda A_\lambda \phi H_\mathrm{u} H_\mathrm{d} + \kappa A_\kappa \phi N^2 + h.c. \nonumber \\ 
  &  + m_0^2(|H_\mathrm{u}|^2 + |H_\mathrm{d}|^2 + |N|^2) + m_\phi^2|\phi|^2 \;.
\end{align}
$\phi$ and $N$ are, respectively, the inflaton
and waterfall fields. 
These fields are singlets of the Minimal Supersymmetric
Standard Model (MSSM) \cite{Chung:2003fi} gauge group and the other
fields are just the usual quarks and Higgs multiplets of the MSSM
with standard MSSM quantum numbers.

This model utilises hybrid inflation 
\cite{Linde:1990gz,Linde:1991km,Copeland:1994vg,Lyth:1996we,Linde:1997sj}  
to provide a simultaneous solution to the strong CP and $\mu$ 
\cite{Chung:2003fi} problems, as we will now briefly outline 
(for a more detailed discussion  
see \cite{Eyton-Williams:2004bm} and the references
therein): 

\begin{itemize}
\item During inflation, the scalar potential reduces to the simple form 
\begin{eqnarray}
V = V(0) - m_\phi^2 \phi^2 \; ,
\end{eqnarray} 
and the vacuum energy dominates the potential during inflation. Inflation ends
by a second order phase transition if $\phi$ reaches a critical value 
\begin{eqnarray}
\phi_{c}^{\pm} = \frac{A_k}{4 k} \left(- 1 \pm \sqrt{1 - 4 \frac{m_N^2}{A_k^2}} 
\right)  .
\end{eqnarray}
In the minimum of the potential after inflation, 
the field values are \cite{Eyton-Williams:2004bm}
\begin{eqnarray}
\< \phi \> &=& - \frac{A_\lambda}{4 \lambda} \; ,\\
\< N \> &=& \frac{A_\lambda}{2 \sqrt{2} \lambda} \sqrt{1 - 4
\frac{m_0^2}{A_\lambda^2}} \; .
\end{eqnarray}
 
\item The model has the Peccei-Quinn (PQ) symmetry U(1)$_{\mathrm{PQ}}$ where
the global charges of the fields satisfy
\begin{eqnarray}
Q_\phi + Q_{H_\mathrm{u}} + Q_{H_\mathrm{d}} = 0 \; , \;\; 
Q_\phi + 2 Q_N = 0 \; .
\end{eqnarray}

\item After inflation has ended the VEV 
of the inflaton $\<\phi\>$ both generates the $\mu$ term (in a similar 
way to the NMSSM
 \cite{NMSSM, NMSSMphenom}), when $\lambda \phi H_\mathrm{u} H_\mathrm{d} \rightarrow
 \mu H_\mathrm{u} H_\mathrm{d}$ and breaks the U(1)$_{\mathrm{PQ}}$ \cite{Peccei:1977hh} 
 symmetry solving the strong CP problem.
\end{itemize}

The model leads to the following constraints on the scales 
and couplings of the
theory: With the VEV of $\phi$ given by $\<\phi\>=-\frac{A_\lambda}{4\lambda}$, 
we can use the constraints on the axion decay constant to determine the value of 
$\lambda$. For soft terms at the TeV scale, we see that an axion decay 
constant should lie in the range $10^{10}\mbox{ GeV}\le f_a \le 10^{13}
\mbox{ GeV}$ (see \cite{Raffelt:1999tx, Turner:1985si} for derivation of the 
allowed region) requires that $\lambda$ lie in the range $10^{-7} \ge \lambda \ge 10^{-10}$.  
If we take the smallest value in this range this allows\footnote{The origin 
of this small coupling is discussed  
in Appendix \ref{sec:Couplings}} 
$\lambda\sim 10^{-10}$ and it is this small coupling that we have used 
in the neutrino sector. 

Requiring the stability of the vacuum post inflation and that inflation ends 
imposes constraints on the range of allowed ratios of the soft terms:
\begin{align}
  8m_0^2>|A_\lambda|^2>4m_0^2 \; . \label{eq:bounds}
\end{align}
Hence we were able to show that:
\begin{align}
  \mu^2=(0.25-0.5)m_0^2\;,
\end{align}
where $m_0$ is a soft scalar mass common to many of the matter fields of 
order a TeV,
whose universality was shown to result from the string
construction.

\section{Conclusions}\label{sec:Conclusions}
We have shown that Type I string theory provides a natural framework for the
 construction of Dirac neutrino models. Small couplings are essential to our
 model and Type I string theory, compactified on an orbifold, is equipped with a
 geometric explanation for these small couplings.  The large ratios between the
 different couplings are explained in terms of the anisotropy of the compact
 space.  These extra dimensions are not especially large since two radii are of
 order the Planck length and one approximately $10^{-8}$ GeV$^{-1}$.  

These small Yukawa couplings allow us to relate physics at very different
scales: namely the neutrino mass, electroweak and Peccei-Quinn scales.  In so
doing we connect the physics of neutrino mass and inflation.  Specifically we
construct a model which consistently describes neutrino mass generation and an
inflationary solution to the strong CP and $\mu$ problems and has its roots in
Type I string theory.  Consistency with the measured values of the MNS matrix is
achieved by generating the mass matrix from non-renormalisable operators.  These
operators are field theoretic in origin, coming from a supersymmetric
generalisation of the Froggatt-Nielsen mechanism.  The set of allowed operators
is restricted by the inclusion of an additional
U(1)$_\mathrm{R}\times\mathbb{Z}_3$ symmetry which leads to the angles and
masses shown in Eq.(\ref{eq:tantheta23}-\ref{eq:m1}).
It would be interesting to extend this model to include 
the quark sector. In such an approach it may be possible to
relate the Cabibbo angle $\theta_\mathrm{C}$ to the neutrino mass
hierarchy $m_2/m_3$ in terms of an expansion parameter $\lambda =
\theta_\mathrm{C}$ in a more direct way than in the see-saw mechanism.

The Dirac nature of neutrino masses would have far reaching 
phenomenological consequences: Dirac neutrinos would, for instance,   
not induce any signal in neutrinoless double beta (0$\nu\beta\beta$) decay
experiments \cite{Aalseth:2004hb}. Since in our scenario the Dirac mass 
matrix is not directly related
to the mass matrices of quarks and charged leptons, there is {\em a priori} no reason
for favouring a hierarchical neutrino mass spectrum compared to an inverted or
quasi-degenerate one. Although we have discussed the case of a hierarchical
spectrum in this work, from a model building point of view a
quasi-degenerate neutrino mass spectrum for Dirac neutrinos can emerge, e.g.,  
from additional Abelian, non-Abelian or discrete symmetries in our scenario. 
Such quasi-degenerate Dirac neutrino masses could be observable in 
Tritium $\beta$-decay experiments like KATRIN \cite{KATRIN}.  
Together with non-observation of 0$\nu\beta\beta$ decay, this could 
prove the Dirac nature of neutrinos.   
In our scenario, the inflaton can only have small couplings of order
$g_{5_1} \sim 10^{-10}$ to matter. This generically implies a low  
reheat temperature after inflation ${\cal O}(1 - 10)$ GeV, 
which is interesting with respect to gravitino (and other similar) 
constraints in some supergravity models \cite{gravitinoproblem}. 
On the other hand, with Dirac neutrinos the original leptogenesis mechanism 
\cite{Fukugita:1986hr} via the out-of equilibrium decay of heavy right-handed 
neutrinos is obviously not available. 
However other versions of leptogenesis, such as
Dirac leptogenesis \cite{Dick:1999je}, which rely on  
sphaleron transitions for converting lepton into baryon asymmetry could
work for a low reheat temperature \cite{Bastero-Gil:2000jd} due to preheating.
In summary, we feel that in addition to interesting theoretical issues
regarding, e.g., the embedding in string theory,  
scenarios of Dirac neutrinos and inflation like the one discussed in this 
work have a rich phenomenology, which deserves further exploration.

\section*{Acknowledgements}
We acknowledge support from the PPARC grant PPA/G/O/2002/00468.

\newpage

\section*{Appendix}
\appendix

\renewcommand{\thesection}{\Alph{section}}
\renewcommand{\thesubsection}{\Alph{section}.\arabic{subsection}}
\def\theequation{\Alph{section}.\arabic{equation}}
\renewcommand{\thetable}{\arabic{table}}
\renewcommand{\thefigure}{\arabic{figure}}
\setcounter{section}{0}
\setcounter{equation}{0}

\section{String Selection Rules} \label{sec:Couplings}
We will now review the properties of Type I string theory relevant for model
building, first presented in \cite{Ibanez:1998rf}, and summarise the model
presented in \cite{Eyton-Williams:2005bg}. We will be working with a D-brane
setup which includes a geometric mechanism for generating small gauge and Yukawa
couplings.   We consider the class of spaces known as orientifolds (see
\cite{Aldazabal:1998mr} for a study of possible orientifolds) requiring the
addition of intersecting stacks of orthogonal D5-branes and space-filling
D9-branes for consistency.  These spaces are all constructed from a 6-torus and
it is the volume and anisotropy of this torus that leads to the generation of a
hierarchy of couplings. The 6-torus itself is constructed out of three 2-tori
each of which has one radius associated with it. We will show that if one radius
is of order $10^{-8}$ $\mbox{GeV}^{-1}$ and the other two are of order
$10^{-18}$ $\mbox{GeV}^{-1}$ then we obtain a coupling of order $10^{-10}$.

After compactification we end up with, in the most general case, a model consisting of three orthogonal stacks of D5-branes and a stack of D9-branes.

The gauge couplings on the branes can be shown to be the following functions of the extra-dimensional radii.
\begin{align}
  g_{5_i}^2=\frac{2\pi \lambda_{\mathrm{I}}}{R_i^2 M_*^2} \label{eq:g5i}
\end{align}
\begin{align}
  g_{9}^2=\frac{2\pi \lambda_{\mathrm{I}}}{R_1^2 R_2^2 R_3^3 M_*^2} \label{eq:g9}
\end{align} 
Where $\lambda_{\mathrm{I}}$ is the ten dimensional dilaton that governs the strength of string interactions, $M_*$ is the Type I string scale and $R_i$ are the radii of compactification.

The non-canonical, $D=4$, $\mathcal{N}=1$ effective superpotential has only
$\mathcal{O}(1)$ Yukawa couplings \cite{Everett:2000up}, but the K{\"a}hler metric,
although diagonal, is significantly different from the identity. To understand
our theory in the low energy, after the dilaton and moduli have acquired vevs, we must canonically normalise the K{\"a}hler potential and take the flat limit in which $M_p \rightarrow \infty$ while $m_{3/2}$ is kept constant \cite{Brignole:1997dp, Kaplunovsky:1993rd}.  This gives a theory containing superfields with canonical kinetic terms interacting via renormalisable operators.  Notice that the Yukawa couplings can be identified with the gauge couplings (up to the $\mathcal{O}(1)$ factors present before normalisation):
\begin{align}
 W = g_9 & \left(C_1^9 C_2^9 C_3^9 + C^{5_1 5_2} C^{5_2 5_3} C^{5_3 5_1}+
  \sum_{i=1}^3 C_i^9 C^{95_i}C^{95_i} \right)
  +\sum_{i,j,k=1}^3g_{5_i}\left(C_1^{5_i}C_2^{5_i}C_3^{5_i}
  \right. \nonumber\\
 &  +  C_i^{5_i}C^{95_i}C^{95_i} + d_{ijk}C^{5_i}_j C^{5_i 5_k}
 C^{5_i 5_k} + \frac{1}{2}d_{ijk}C^{5_j 5_k} C^{95_j} C^{95_k})
 . \label{eq:WGen}
\end{align}
where the $C$ terms are low energy excitations of strings: 
charged chiral superfields.  The superscripts denote the branes 
which the strings end on and terms with different subscripts 
transform differently under the gauge group associated with the brane.

The $D=4$ Planck scale is related to the string scale by
\begin{align}
  M_p^2=\frac{8M_*^8 R_1^2 R_2^2 R_3^2}{\lambda_{\mathrm{I}}^2}.
\end{align}
From this and Eqs.~(\ref{eq:g5i}) to (\ref{eq:g9}) we find that
\begin{align}
  g_{5_1}g_{5_2}g_{5_3}g_{9}=32\pi^2\left(\frac{M_*}{M_p}\right)^2. \label{eq:gaugereln}
\end{align}
Another important relation is
\begin{align}
  \lambda_{\mathrm{I}} = \frac{g_{5_1} g_{5_2} g_{5_3}}{2\pi g_9}\; . \label{eq:lambdaI}
\end{align}
In the model discussed in the next section, we
shall require at least one coupling of $\mathcal{O}(10^{-10})$ and one
of $\mathcal{O}(1)$.  According to the above results, this constrains
the size of our radii and the value of the string scale.  For
definiteness we consider the case where $g_{5_1}\sim 10^{-10}$ and the
remaining gauge couplings are all $\mathcal{O}(1)$. From
Eq.~(\ref{eq:gaugereln}) we see this is clearly allowed if we have a
$10^{13}$ GeV string scale. Specifically our couplings are
$g_{5_2}=\sqrt{\frac{4\pi}{24}}$ (to give $\alpha_{\mbox{\tiny
GUT}}=1/24$, consistent with gauge coupling unification),
$g_{5_3}=g_9=2$ and $g_{5_1}=10^{-10}$ gives $M_*= 10^{13}$ GeV.

The hierarchy in gauge couplings corresponds to a hierarchy in the radii.  Using Eqs.~(\ref{eq:lambdaI}) and (\ref{eq:g5i}) for the above couplings we find that
\begin{align}
  R_1^{-1} = 1.3 \times 10^8 \mbox{ GeV} \\
  R_2^{-1} = 9.1 \times 10^{17} \mbox{ GeV} \label{eq:R2} \\
  R_3^{-1} = 2.4 \times 10^{18} \mbox{ GeV}.
\end{align}
These radii are all too small to have Kaluza-Klein (KK) or winding modes that will be readily excitable at collider energies. The winding modes of $R_1$ are $\approx n 10^{18}$ GeV and $R_2$ and $R_3$ have winding modes of $\approx n 10^{8}$ GeV.  The KK modes for $R_1$ are $\approx n 10^8$ GeV and $R_2$ and $R_3$ are $\approx n 10^{18}$ GeV. In principle these massive modes could affect inflation.  However the inflationary scale is $10^8$ GeV so it is unlikely that these modes would appear with any great abundance.

\subsection{Methodology}

Our approach in this paper is one of string inspired phenomenology.  We make use of a number of the generic properties of low energy effective string theory so as to keep our analysis as general as possible and avoid specialising to a particular model.  The rules we enforce are:
\begin{itemize}
\item All supersymmetric terms must be found within the low energy effective
superpotential Eq.~(\ref{eq:Wintro}).
\item The string states, $C^{5_1 5_2}$ etc., can represent more than one low
energy field.
\item Each low energy field can only be assigned to one string state.
\item The gauge quantum numbers of a string state are determined by the stacks
of branes which it ends on.
\end{itemize}
We will now clarify and expand on the final point.  Clearly the ends of the string can either both be on the same stack of branes or attached to two different stacks.  The string ends' locations determine the transformation properties of their low energy excitations since the brane stacks have an associated gauge group under which the strings transform. If both string ends attach to the same brane stack then it is commonly the case that fields transform as reducible representations of that stack's gauge group, typically U(N).  In the other case, with ends on different stacks, then the fields generally transform as fundamental representations of both stack's gauge groups.  We impose this requirement in our model building.

\section{Our Convention for Lepton Mixing}\label{conventions}
For the mass matrix of the charged leptons 
$m^{\mathrm{E}}_{\mathrm{LR}}=Y_\mathrm{e} v_\mathrm{d}$ defined by
$\mathcal{L}_\mathrm{e}=-m^{\mathrm{E}}_{\mathrm{LR}} \overline e^f_{\mathrm{L}}
e^{\mathrm{c}f}_{\mathrm{R}}$ + h.c.
and for the Dirac neutrino mass
matrix $m^\nu_{\mathrm{LR}}=Y_\nu v_\mathrm{u}$ defined by
$\mathcal{L}_\nu=-m^\nu_{\mathrm{LR}} \overline \nu^f_{\mathrm{L}}
\nu^{\mathrm{c}f}_{\mathrm{R}}$ 
+ h.c., where
$v_\mathrm{u} = \< H^0_\mathrm{u}\>$ and 
$v_\mathrm{d} = \< H^0_\mathrm{d}\>$, the change from flavour basis to mass
eigenbasis can be performed with the unitary diagonalization matrices
$U_{e_\mathrm{L}},U_{e_\mathrm{R}}$ and
$U_{\nu_\mathrm{L}},U_{\nu_\mathrm{R}}$ by
\begin{eqnarray}\label{eq:DiagMe}
U_{e_\mathrm{L}} \, m^{\mathrm{E}}_{\mathrm{LR}} \,U^\dagger_{e_\mathrm{R}} =
\left(\begin{array}{ccc}
\!m_e&0&0\!\\
\!0&m_\mu&0\!\\
\!0&0&m_\tau\!
\end{array}
\right)\! , \quad
U_{\nu_\mathrm{L}} \, m^\nu_{\mathrm{LR}} \,U^\dagger_{\nu_\mathrm{R}} =
\left(\begin{array}{ccc}
\!m_1&0&0\!\\
\!0&m_2&0\!\\
\!0&0&m_3\!
\end{array}
\right)\! .
\end{eqnarray}
The mixing matrix in the lepton sector, the MNS matrix, is then given by
\begin{eqnarray}\label{Eq:MNS_Definition}
U_{\mathrm{MNS}} = U_{e_\mathrm{L}} U^\dagger_{\nu_\mathrm{L}}\; .
\end{eqnarray}
We use the parameterisation
$
U_{\mathrm{MNS}} = R_{23} U_{13} R_{12}
$
with $R_{23}, U_{13}, R_{12}$ defined as
\begin{eqnarray}
R_{12}:=
\left(\begin{array}{ccc}
  c_{12} & s_{12} & 0\\
  -s_{12}&c_{12} & 0\\
  0&0&1\end{array}\right)
  , \;\;
U_{13}:=\left(\begin{array}{ccc}
   c_{13} & 0 & \widetilde{s}_{13}\\
  0&1& 0\\
  - \widetilde{s}^{\: *}_{13}&0&c_{13}\end{array}\right)  
  , \;\;
R_{23}:=\left(\begin{array}{ccc}
 1 & 0 & 0\\
0&c_{23} & s_{23}\\
0&-s_{23}&c_{23}
 \end{array}\right), \nonumber 
\end{eqnarray}
and where $s_{ij}$ and $c_{ij}$ stand for $\sin (\theta_{ij})$ and $\cos
(\theta_{ij})$, respectively. 
 $\delta$ is the Dirac CP phase relevant for neutrino oscillations 
 and we have defined $\widetilde{s}_{13}:=s_{13}e^{- i \delta}$.

%%% Correctly ordered bibliography as generated by sort2
%%% (c) I Peddie 2005. This is released under the GNU Public licence
%%% See www.gnu.org for details. 

\end{document}